\documentclass[12pt,preprintnumbers,aps,amssymb,nofootinbib,amsmath]{revtex4}
\usepackage{epsfig}
\usepackage{graphicx}
\usepackage{epstopdf}
\def\eq#1{{(\ref{#1})}}
\def\fig#1{{Fig.~\ref{#1}}}

\newcommand{\beq}{\begin{equation}}
\newcommand{\eeq}{\end{equation}}
\newcommand{\beqar}[1]{\begin{eqnarray}\label{#1}}
\newcommand{\eeqar}{\end{eqnarray}}
\newcommand{\bas}{\bar{\alpha}_s}
\newcommand{\as}{\alpha_s}
\newcommand{\un}{\underline}

\newcommand{\fkn}{\mathfrak{N}}
\begin{document}

\preprint{}

\title {Chaos in the Color Glass Condensate}

\author{Dmitri Kharzeev}

\affiliation{ Nuclear Theory Group, \\ Physics Department, Brookhaven National Laboratory,\\
Upton, NY 11973-5000, USA}

\author{Kirill Tuchin}
 
\affiliation{ Nuclear Theory Group, \\ Physics Department, Brookhaven National Laboratory,\\
Upton, NY 11973-5000, USA}

\preprint{BNL-NT-05/3}
\date{\today}

\begin{abstract} 
The number of gluons in the hadron wave function is discrete, and their 
formation in the chain of small $x$ evolution occurs over discrete rapidity intervals of $\Delta y \simeq 1/\as$. 
We therefore consider the evolution as a discrete quantum process. We show that the discrete version of the 
mean-field Kovchegov evolution equation gives 
rise to strong fluctuations in the scattering amplitude, not present in the continuous equation. 
We find that if the linear evolution is as fast as predicted by the perturbative BFKL dynamics, the scattering amplitude at high energies exhibits a chaotic behavior.
As a consequence, the properties of diffraction at high energies become universal.
\end{abstract}
\maketitle 


The wave function of an ultra--relativistic hadron is described by the  Color Glass Condensate (CGC) \cite{GLR,MQ,MV,jmwlk,ILM} --  a quasi-classical non-Abelian Weizsacker-Williams field \cite{MV,Kov}. It emerges when the occupation number of the bremsstrahlung gluons emitted at a given impact parameter exceeds unity and eventually saturates at $\sim 1/\as$.  It has been argued that in a big nucleus, such that $\as A^{1/3}\gg1$,  and not very high energies the mean-field treatment is a reasonable approximation of the CGC evolution equations. In this approximation the interaction of a dipole with a big nucleus is decribed by so-called fan diagrams \cite{GLR}.  
However in general the quantum fluctuations around the classical solution can strongly modify the scattering amplitudes. 
One way to understand the origin of these fluctuations is to consider the gluon emission as a discrete process \cite{IMM}. It was argued in \cite{IMrare} that a color dipole wave function is dominated by rare dipole  configurations which lead to strong fluctuations of the dipole (gluon) density and, as a consequence, to the fluctuations of the scattering amplitude.  It was proposed to take such fluctuations into account by introducing new terms into the Kovchegov equation \cite{Iancu:2004iy,Mueller:2005ut}. 

In this paper we take a different approach.  An introduction of an infrared cutoff $\Lambda$ on the momentum of the emitted gluons amounts to imposing the boundary condition. This is equivalent to the quantization 
of the gluon modes in a box of size $L \sim \Lambda^{-1}$, in which case the spectrum of the emitted 
gluons and their number become discrete.   
The formation of a gluon occurs over a rapidity interval of $\Delta y \simeq 1/\as$. Therefore, the evolution in rapidity 
can be considered as a discrete quantum process, where each subsequent step occurs when $\Delta y\, \as \simeq 1$.
We will show that the discrete version of the mean-field Balitsky--Kovchegov evolution equation \cite{Balitsky:1995ub,Kovchegov:1999yj} gives 
rise to chaotic behavior of the scattering amplitude. We find that even the event--averaged scattering amplitude differs 
significantly from the continuous result.  
We will argue that this chaotic behavior is a general feature of discrete evolution when the growth 
of the scattering amplitude is sufficiently fast; this appears to be the case for the perturbative BFKL \cite{bfkl} evolution.

In the conventional mean-field approximation Kovchegov equation \cite{Balitsky:1995ub,Kovchegov:1999yj} is formulated for the scattering amplitude $N(\un x_0-\un x_1,\un b, y)$ of the color dipole of transverse size $\un x_0-\un x_1$,  at impact parameter $\un b=(\un x_0+\un x_1)/2$ and rapidity $y$. In this paper we limit ourselves to the case of the most central collisions in which case 
the scattering amplitude can be evaluated at a fixed impact parameter. Hence omitting $\un b$ the Kovchegov equation reads
\beq\label{KovX}
\frac{\partial N(\un x_{01},y)}{\partial y}\,=\,\frac{\bas}{2\pi} \int\,d^2x_2\frac{\un x_{01}^2}{\un x_{12}^2\,\un x_{02}^2}\,
(N(\un x_{12},y)\,+\,N(\un x_{02},y)\,-\,N(\un x_{01},y)\,-\,N(\un x_{12},y)\,N(\un x_{02},y))\,
\eeq
It will be more convenient for our discussion to rewrite this equation in the momentum space. Defining
\beq\label{tilden}
\tilde N(\un k, y)\,=\,\int d^2x\,\frac{1}{\un x^2}\,e^{i\un k\cdot \un x}\,N(\un x,y)
\eeq	
we can write \eq{KovX} as \cite{Kovchegov:1999ua}
\beq\label{KovK}
\frac{\partial\tilde N(\un k,y)}{\partial y}\,=\,\bas\,\chi(\hat\gamma)\,\tilde N(\un k,y)\,-\,\bas\,\tilde N^2(\un k,y)\,,
\eeq
where $\chi(\gamma)$ is the familiar leading eigenvalue of the BFKL equation
\beq\label{chi}
\chi(\gamma)\,=\,2\,\psi(1)\,-\,\psi(1\,-\,\gamma)\,-\,\psi(\gamma)
\eeq
and the anomalous dimension operator is given by
\beq\label{anomd}
\hat\gamma(k)\,=\,1+\,\frac{\partial}{\partial\,\ln k^2}\,.
\eeq

\begin{figure}
\includegraphics[width=10cm]{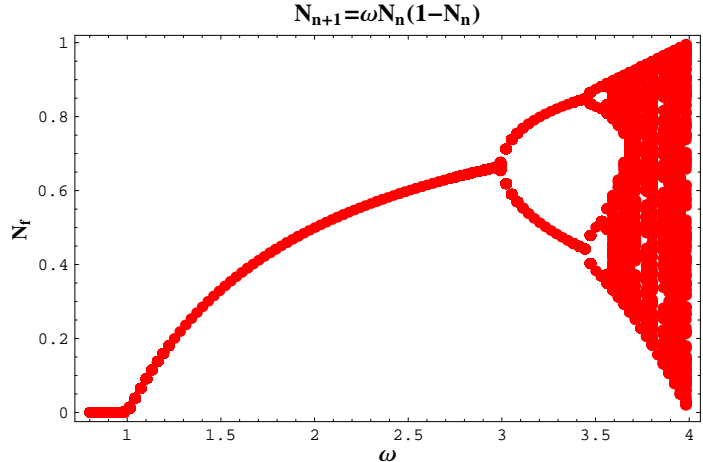}
\caption{Bifurcation map: the fixed points of small $x$ evolution as a function of Malthusian 
parameter $\omega$.}\label{fig:bifurc}
\end{figure}

Eqs.~\eq{KovX} and \eq{KovK} are written for the event averaged scattering amplitude. Let us, in the spirit of Ref.~\cite{IMM}, consider the evolution in a given event.  The emission of one low $x$ gluon is of the order $\bas y\sim 1$ contribution to the scattering amplitude. Therefore, we can enumerate the emitted gluons by discrete values $n$ of the ``time" parameter $\bas y$. Thus the discrete
version of Kovchegov equation takes the form
\beq\label{desK}
\tilde N_{n+1}(\un k,y)\,=\,(1\,+\,\chi(\hat\gamma))\,\tilde N_n(\un k,y)\,- \,\tilde N_n^2(\un k,y)\,,
\eeq
Let us expand $\chi$ near the saddle point $\gamma_0$. In this paper we keep only the first term in the expansion for the sake of simplicity -- this toy model suffices to illustrate our main idea.
Introducing the rescaled amplitude $\mathfrak{N}=\tilde N/(\chi(\gamma_0)\,+\,1)$ we can write the Kovchegov equation as 
\beq\label{resc}
\mathfrak{N}_{n+1}\,=\,(\chi(\gamma_0)\,+\,1)\,\fkn_n\,(1\,-\,\fkn_n)\,.
\eeq
One can recognize \eq{resc} as the famous logistic map \cite{logmap};  $\omega\equiv\chi(\gamma_0)+1$ is called the Malthusian parameter. The properties of its solutions are listed below (see e.\ g.\ \cite{Math}). 
\begin{figure}
\includegraphics[width=8cm]{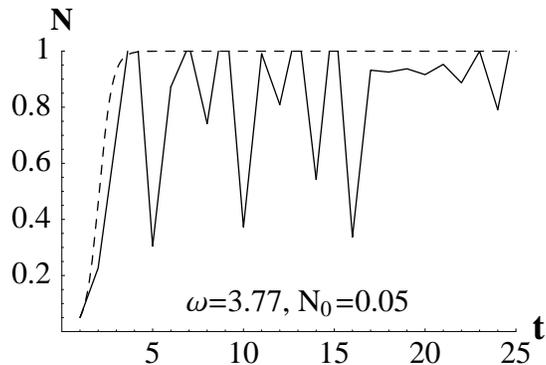}
\caption{The scattering amplitude (scaled by $\omega/(\omega-1)$) as a function of the number of evolution steps. The dashed 
curve indicates the continuous limit; the discrete evolution (solid line) leads to the chaotic behavior.
}\label{fig:chaos}
\end{figure}
\begin{enumerate}

\item $1<\omega<3$:  the scattering amplitude folows the sigmoid curve 
towards the saddle point value $\fkn_f$ (the fixed point).  
Since $\fkn_f<1$, the black disc limit is not achieved even at asymptotically high energies.

\item $3\le\omega<\omega_2$, where $\omega_2\approx 3.449$.  At these values of $\omega$ the scattering amplitude at $n\rightarrow \infty$ does not converge to a single limit -- instead, it oscillates between two fixed points. Mathematically, $\fkn$ develops a pitchfork bifurcation at $\omega=3$ at which point the 2-cycle begins.  As seen in \fig{fig:bifurc} the next bifurcation happens at $\omega_4\approx 3.449$; it starts a 4-cycle.  The 8-cycle starts at $\omega_8\approx 3.56$ and so on. Note, that the behavior of $\fkn$ at $n\rightarrow \infty$ is independent of the inital condition at $n=1$ as long as  $\omega<3.57$.

\item If $\omega\ge \omega_\mathrm{ap}$, where $\omega_\mathrm{ap}\approx 3.57$ is the  accumulation point, periodicity gives way to chaos.  In other words the scattering amplitude
exhibits irregular, unpredictable behavior which manifests itself in a sensitivity to small changes in the initial condition. Recall that  the BFKL saddle point is $\gamma_\mathrm{BFKL}=1/2$ at which $\chi(\gamma_\mathrm{BFKL})=4\ln 2\approx 2.77$, thus $\omega_\mathrm{BFKL}=3.77$.  In perturbation theory $\omega$ cannot take values less than $\omega_\mathrm{BFKL}$. Therefore, we conclude that 
the perturbative high energy evolution is chaotic. 

\end{enumerate}

By averaging over all events one can define the mean value of the scattering amplitude.
However, this procedure hides a lot of interesting physics. The most obvious example of this is diffraction, which measures 
the strength of fluctuations in the inelastic cross section \cite{diff}. The arguments given in points 2--3 above show that diffraction is a significant part of the total inelastic cross section at very high energies, and is universal (independent from the 
properties of the target). 

It is important to emphasize, that our model treats the high energy evolution process classically. We neglected the fact that the gluon emission is a stochastic process. The emission ``time", i.e.\ the rapidity interval $dy$ over which a gluon is radiated, varies from event to event. In other words, gluon emission is a quantum process which may or may not occur with a certain probability once the rapidity interval of the collision $y$ is increased by $dy$. Full treatment of the discrete BK equation requires taking these effects into account. However, unfortunately BK equation is known to resist all attempts of analytical solution, and our hope at present is to develop a meaningful approximation. Thus, in our paper we suggested an approximation in which the gluons are emitted over a fixed ``time" defined by $\bar\alpha_sdy =C$ with $C=1$.
To justify this assumption, let us note that BFKL takes into account only fast gluons, i.e.\  those with $C\sim 1$. It is beyond the leading logarithmic (LL) approximation to take into account slow gluons. Moreover, it is known that an account of NLL corrections effectively leads to imposing a rapidity veto \cite{veto} on the emission of gluons with close rapidities, which restricts production of gluons with small $C$ (this is due to an effective repulsion between the emitted gluons induced at the NLL level). Therefore $C$ is bounded from below by a number close to one. On the other hand the probability that no gluon is emitted when $C$ becomes larger than one is very small if we choose the high density initial condition, such as the one given by the McLerran-Venugopalan model\cite{MV}.  Therefore, $C$ takes random values around 1, but the effective dispersion can 
be expected quite small.  

We realize that  the quantum fluctuations can affect the effective value of $C$ and push the onset of chaos to a different kinematic region, but we believe that this effect is not going to be dramatic. The present paper is only the first step in the exploration of the discrete BK equation. Our aim is to stress its highly nontrivial structure which might have an important impact on the high energy theory and phenomenology. The question about the stability of the found peculiar classical solution with respect to the stochastic quantum fluctuations is very important and we are going to address it in the forthcoming work.

\vspace{0.3cm}

The model used in this letter is admittedly oversimplified: we neglected the diffusion in transverse momentum, stochasticity of gluon emission and the dynamical fluctuations beyond the mean field approximation. Nevertheless, we hope that at least some of the features of discrete quantum evolution at small $x$ will survive a more realistic treatment. The chaotic features of small $x$ evolution 
open a new intriguing prospective on the studies of hadron and nuclear interactions at high energies.


\acknowledgments

The authors would like to thank Yuri Kovchegov, Alex Kovner, Misha Kozlov, Genya Levin, Larry McLerran, Al Mueller, Anna Stasto for informative and helpful discussions and comments.
This research was supported by the U.S. Department of
Energy under Grant No. DE-AC02-98CH10886.


\end{document}